\documentclass{article}
\newcommand{\bfr}{\begin{flushright}}
\newcommand{\efr}{\end{flushright}}
 
\begin{document}
\title{Wilson-Loop Symmetry Breaking Reexamined
}
\author{ 
Atsushi Nakamula\\
Department of Physics, Tokyo Metropolitan University,\\
Setagaya-ku, Tokyo 158, Japan\\
and\\
Kiyoshi Shiraishi\\
Institute for Nuclear Study, University of Tokyo, \\
Midori-cho, Tanashi,
Tokyo 188, Japan
}
\date{Phys. Lett. {\bf B215}, No.~3 (1988) 551; 
{\bf B218}, No.~4 (1989) 508 (E).
}
\maketitle
\begin{abstract}
The splitting in energy of gauge field vacua on the non-simply connected
space $S^3/Z_2$ is reconsidered. We show the calculation to the one-loop
level for a Yang-Mills vector with a ghost field. We confirm our
previous result and give a solution to the question posed by Freire,
Rom\~ao and Barroso. 
\end{abstract}

The origin of the gauge symmetry group, which
includes that of the standard model, is still unclear. Unified theories
of fundamental forces assume a large symmetry group which is to be
broken at the low-energy scale.

Recently developments of string theories \cite{1} originated from the
fact that the gauge group is uniquely determined due to
quantum-mechanical consistency \cite{2}. The theories are formulated in
higher dimensions. In more recent days, a large number of
``four-dimensional string models'' \cite{3} with various symmetry groups
were constructed at the sacrifice of uniqueness; the ``right'' choice of
gauge group for string theory is still unknown, however \cite{4}.

There is the interesting possibility that compactification and gauge
symmetry breaking take place altogether in higher-dimensional theory
with the gauge theory of a large group. The mechanism dubbed the
Wilson-loop mechanism has been used in the context
of gauge symmetry breaking in superstring theories \cite{5}.

But the phenomenon can be completely explained
by means of the field-theoretical concept. Indeed,
simple examples with an extra one-dimensional compact space, a circle,
were provided by Hosotani \cite{6}. He also pointed out the importance of
the calculation of a Casimir-like energy, in order to choose a
``correct'' vacuum; namely, we have to look for the gauge vacuum which
has the lowest energy. 

A more complex model was first considered by
Evans and Ovrut \cite{7}. The model is formulated in the
spacetime $M_4\times S^3/Z_2$, where $M_4$ denotes flat four-dimensional
spacetime and $S^3/Z_2$ means a three-sphere on which antipodal points
are identified. They, however, did not give an explicit computation of
the value of vacuum energies.

One of the present authors carried out \cite{8} an evaluation of the
one-loop vacuum energy difference by a sort of dimensional
regularization a la Candelas and Weinberg \cite{9}. Then he concluded
that the lower energy vacuum is the vacuum which realizes the unbroken
gauge symmetry. 

Recently, however, the authors of ref.~\cite{10} claim
that the vacuum of broken gauge symmetry has a
lower energy than that of unbroken symmetry.

We examine in this letter the one-loop calculation
of the energy difference of two vacua which arise in
the model on the spacetime $M_4\times S^3/Z_2$, and give a
solution to the question. Most of our results were
published in ref.~\cite{8}. We carefully confirm the results as one will
see. 

Let the metric on $S^3$ be of the form
\begin{equation}
ds^2 = r^2 [d\psi^2 + \sin^2\psi\,(d^2\theta + \sin^2\theta d\phi^2 )]\,,
\end{equation}
where $r$ is the radius of $S^3$. A $Z_2$ transformation is
defined by
\begin{equation}
\psi\rightarrow\pi-\psi\,,\quad\theta\rightarrow\pi-
\theta\,,\quad\phi\rightarrow\phi+\pi\,.
\end{equation}
If we pick out invariant states or harmonics under
this transformation, identification of the antipodal
points on $S^3$ is attained \cite{7,8,10}.

Throughout this letter, the background spacetime
is fixed (by hand). Furthermore, in our terminology,
the words ``energy'' and ``energy density'' are used
with no strict distinction. There appears no ambiguity or confusion as
long as we consider a static, fixed spacetime.

As a simple, specific model, and for the sake of
comparison, we suppose an $SU(3)$ gauge field only,
although the choice of gauge group is irrelevant to the
conclusion, as we will see.

Here, capital Latin indices have the range $M,
N=0,..., 6$; Greek indices $\mu, \nu$ run over $0,1,2,3$ and
lowercase Latin indices $m, n$ run over $4,5,6$. Thus the
$SU(3)$ gauge field is written as
\begin{equation}
A^a_M = (A^a_\mu, A^a_m)\,,
\end{equation}
where $a$ is the group index of the adjoint representation.
We have to note, first of all, that the classical configurations of the
gauge field allowed on $S^3/Z_2$ are not only trivial ones like
\begin{equation}
A^a_m{}^{(1)}= (0, 0, 0)\, ,
\end{equation}
but also, for example \cite{7},
\begin{equation}
A^a_m{}^{(2)}= (0, 0, 2\sqrt{3}\delta^{a8})\, .
\end{equation}
or \cite{10}
\begin{equation}
A^a_m{}^{(3)}= (0, 0, 2\delta^{a3})\, ,
\end{equation}
to satisfy $F_{mn}^a=D_mA_n^a-D_nA_m^a+f^{abc}A_m^bA_n^c=0$,
where $D_m$ is the covariant derivative and $f^{abc}$ is the
structure constant of the group. These configurations
of which the field strengths vanish are called vacuum
gauge fields. The above gauge configurations have $Z_2$
invariance defined on $S^3$. The configurations $A^a_m{}^{(1)}$
and $A^a_m{}^{(2)}$ are inequivalent modulo a proper gauge
transformation on $S^3/Z_2$; but $A^a_m{}^{(2)}$ and $A^a_m{}^{(3)}$ are
connected by a global group rotation. Thus the vacuum associated with
$A^a_m{}^{(2)}$ and that of $A^a_m{}^{(3)}$ are identical. In
phenomenology one can think of the vacuum gauge field as a configuration
of adjoint Higgs fields.

Now, we are going to calculate the energy difference between the
different gauge vacua. To compute the one-loop vacuum energy, we only
need to know eigenstates and eigenvalues of Laplacians which function
on the quantum fluctuation of the fields \cite{9}. The eigenstates of
Laplacians defined on $S^3$ are classified even or odd under the $Z_2$
transformation (2)
\cite{7,8,10}. Namely, the harmonics of the even modes
are invariant under the $Z_2$ transformation while those
of the odd modes change sign under the $Z_2$
transformation.

In the trivial vacuum ($A^a_m{}^{(1)}$) on $S^3/Z_2$ all states
are even modes, on the other hand, in the nontrivial
vacua ($A^a_m{}^{(2)}$) on $S^3/Z_2$ the states corresponding to
the component which do not commute with the
vacuum gauge fields are odd modes because of their
coupling to the vacuum field; the other states corresponding to the
commuting component remain to be even modes. In general, the eigenvalues
and eigenfunctions of Laplacians on $S^N$ ($N$-sphere) for various
tensor fields are well known \cite{11} and ``even'' and
``odd'' modes are apparently distinguished from each
other by a quantum number which also directly determines the
eigenvalue \cite{7,8,10}. 

Of course, the zero mode of the gauge field
belongs to the even modes. Thus the gauge symmetry is reduced to be
small in nontrivial vacua since the odd modes involve massive
excitations only. 

Obviously, the total difference in energy is
proportional to the difference in energy for a single degree of freedom;
we do not need group theoretical complexity but need only a naive
counting of the number of group indices attached to the components that
are noncommuting with the vacuum gauge fields (after
diagonalization, if necessary).

Let us denote the energy (density) of each vacuum
as $E^{(1)}$ and $E^{(2)}$. Further, we define the energy difference
for a single degree of the group index, $\Delta E$, which is the
difference in the vacuum energy obtained from the one-loop calculation
adopting even and odd modes on $S^3$ \cite{7,8,10} (see further on).
Because it is $Z_2$ that classifies the harmonics on $S^3$, the energy
difference between any vacua is determined by the single quantity
$\Delta E$. The difference in energy is expressed as 
\begin{equation}
E^{(1)} - E^{(2)} =4\Delta E\,, 
\label{eq5}
\end{equation}
because the vacuum of the gauge configuration
$A^{(2)}_m$ realizes four massless gauge bosons for
$SU(2)\times U(1)$. Thus the sign of $\Delta E$ is only relevant
to the choice of the lower energy vacuum, that is to
say, whether gauge symmetry breaking occurs or not.

Next we will give the derivation of $\Delta E$.

A higher dimensional vector field contains many
kinds of fields, or harmonics, from the point of view
of four dimensions after dimensional reduction. The
most naive classification divides $A^a_M$ into $A^a_\mu$, and
$A^a_m$. The eigenfunctions of the Laplacian for $A^a_\mu$ must
contain scalar harmonics on $S^3/Z_2$, in the usual sense
of Kaluza-Klein theory. Thus the eigenvalue is the
same as that of a scalar field on $S^3/Z_2$ and the degeneracy is four
times that since the flat spacetime index runs from $0$ to $3$.

The eigenfunctions for $A^a_m$ have already been investigated \cite{11}.
There are two kinds of eigenfunctions. One is the transverse part, the
other is the longitudinal part; the latter is known to be expressed
as a covariant derivative of the scalar harmonics on
$S^3$ \cite{11}; and the eigenvalues of the longitudinal vector
including the contribution of the Ricci curvature tensor also coincide
with the ones of the scalar harmonics.
The even and odd modes under $Z_2$ transformation for the
longitudinal vector harmonics are similarly defined as the scalar case.

Finally, we consider a ghost
field. A complex ghost field for vector fields has two scalar degrees of
freedom and a negative contribution to the one-loop energy.

We define the energy differences per degree of freedom for a scalar, a
transverse vector, a longitudinal vector on $S^3/Z_2$, a complex ghost
and a four-dimensional vector as $\Delta E_S$, $\Delta E_{TV}$,
$\Delta E_{LV}$, $\Delta E_G$ and
$\Delta E(A^a_\mu)$. For example, $\Delta E$ is formally expressed as an
infinite sum a la Candelas and Weinberg \cite{9};
\begin{equation}
\Delta E_S = \lim_{n\rightarrow
4}\left(-\frac{1}{2(4\pi)^{n/2}r^n}\Gamma\left(-\frac{n}{2}\right)
\sum_{l=0}^\infty(-1)^l(l+1)^2[l(l+2)]^{n/2}\right)
\label{eq6}
\end{equation}

This is just the difference between the even and odd
sectors of the spectra on $S^3$\cite{8,10}. Each sector corresponds to
even $l$ and odd $l$ respectively. The other $\Delta E$'s can be written
in the same manner \cite{12}.

From the above consideration, there are the following relations among
the $\Delta E$'s:
\begin{equation}
\Delta E_{LV}=\Delta E_S\,,\quad \Delta E_G = - 2\Delta E_S\,,
\quad \Delta E(A^a_\mu) = 4\Delta E_S\,.
\label{eq7}
\end{equation}

The total energy difference $\Delta E$ for a vector field in
$M_4 \times S^3/Z_2$ is written as
\begin{eqnarray}
\Delta E&=& \Delta E(A^a_\mu)+\Delta E_{TV}+\Delta E_{LV}+ \Delta E_G
\nonumber \\
&=& 4\Delta E_S+\Delta E_{TV}+\Delta E_S-2\Delta E_S\nonumber \\
&=&\Delta E_{TV}+3\Delta E_S\,.
\label{eq8}
\end{eqnarray}

The authors of ref.~\cite{11} calculated $\Delta E_{TV}$ and $\Delta
E_G$, however they seemed to forget not only the longitudinal vector but
also the contribution from a four-dimensional vector. The values of
$\Delta E_{TV}$ and $\Delta E_S$ can be read from ref.~\cite{8}, they
are
\begin{equation}
\Delta E_{TV}=6.312\times 10^{-3} r^{-4}\, ,
\label{eq9a}
\end{equation}
\begin{equation}
\Delta E_S=-5.116\times 10^{-3} r^{-4}\, . 
\label{eq9b}
\end{equation}

Accordingly $\Delta E$ [eq.~(\ref{eq8})] is negative, 
thus an
unbroken gauge vacuum is favorable. This conclusion is also applicable
to any gauge group as long as we consider $S^3/Z_2$. 

The authors of
ref.~\cite{7} omitted a concrete calculation o n $M_4\times S^3/Z_2$ .
The authors of ref.~\cite{11} drew the wrong conclusion because they
forgot to consider the contribution of the four-dimensional vectors
$A^a_\mu$. The confirmation of the counting of degrees of freedom can be
done by computing the free energy that concerns each vacuum by means of
the introduction of temperature.

We now consider the one-loop correction to the
``potential'' obtained in ref.~\cite{13}. The authors of ref.~\cite{13}
discovered the path that connects two gauge vacua. We would like to
consider the loop correction to the potential in order to study the
(quasi-) stable domain wall structure at zero and finite
temperature \cite{12}.

\bigskip

K.S. is grateful to the Japan Society for the
Promotion of Science for a fellowship. He also thanks
Iwanami F\=ujukai for financial aid.

The author would like to thank J.S. Dowker and
S. P. Jadhav for pointing out the error in the paper.


\end{document}